\documentclass[useAMS,usenatbib,usegraphicx]{mn2e}
\usepackage{pdflscape}
\usepackage{lscape}
\usepackage{amssymb}
\usepackage{longtable}
\usepackage{endnotes}
\bibpunct{(}{)}{;}{a}{,}{,}

\usepackage{graphicx}
\usepackage{helvet}
\usepackage{natbib}
\usepackage{setspace}
\def\apjl{{ApJ}}%
\def\apjs{{ApJS}}%
\def\aap{{A\&A}}%
\title[Supernova 2011fu]{Light curve and spectral evolution of the Type IIb SN 2011fu}
\author[Brajesh Kumar et al.]{Brajesh Kumar$^{1,2}$ \thanks{E-mail:
brajesh@aries.res.in, brajesharies@gmail.com}, 
S. B. Pandey$^{1}$, D. K. Sahu$^{3}$, J. Vinko$^{4}$, 
\newauthor A. S. Moskvitin$^{5}$, G. C. Anupama$^{3}$, V. K. Bhatt$^{1}$, A. Ordasi$^{6}$,
\newauthor A. Nagy$^{4}$, V. V. Sokolov$^{5}$, T. N. Sokolova$^{5}$, V. N. Komarova$^{5}$,
\newauthor Brijesh Kumar$^{1}$, Subhash Bose$^{1}$, Rupak Roy$^{1}$, Ram Sagar$^{1}$\\
\\
$^{1}$Aryabhatta Research Institute of observational sciencES, Manora Peak,
Nainital 263129, India\\
$^{2}$Institut d'Astrophysique et de G\'{e}ophysique, Universit\'{e} de
Li\`{e}ge, All\'{e}e du 6 Ao\^{u}t 17, B\^{a}t B5c,
      4000 Li\`{e}ge, Belgium \\
$^{3}$Indian Institute of Astrophysics, Koramangala, Bangalore 560 034, India\\
$^{4}$Department of Optics \& Quantum Electronics, University of Szeged, D\'om
t\'er 9, Szeged, Hungary\\
$^{5}$Special Astrophysical Observatory, Nizhnij Arkhyz, Karachaevo-Cherkesia,
369167 Russia\\
$^{6}$Department of Experimental Physics, University of Szeged, D\'om t\'er 9,
Szeged, Hungary\\
} 
      
\begin{document}

\date{Accepted ------------, Received ------------; in original form ------------}
\pagerange{\pageref{firstpage}--\pageref{lastpage}} \pubyear{}
\maketitle
\label{firstpage}
\begin{abstract}
We present the low-resolution spectroscopic and \emph{UBVRI} broad-band photometric 
investigations of the Type IIb supernova 2011fu, discovered in UGC~01626. The photometric 
follow-up of this event has been initiated a few days after the explosion and covers 
a period of about 175 days. The early-phase light curve shows a rise followed by steep decay 
in all bands and shares properties very similar to that seen in case of SN~1993J, 
with a possible detection of the adiabatic cooling phase. Modelling of 
the quasi-bolometric light curve suggests that the progenitor had an extended 
($\sim 1 \times 10^{13}$ cm), low-mass 
($\sim 0.1$ $M_\odot$) H-rich envelope on top of a dense, compact ($\sim 2 \times 10^{11}$ cm), more massive 
($\sim$ 1.1 $M_\odot$) He-rich core. The nickel mass synthesized during the explosion
was found to be $\sim$ 0.21 $M_\odot$, slightly larger than seen in case of other Type IIb SNe. 
The spectral modelling performed with \texttt{SYNOW} suggests that the early-phase line 
velocities for H and Fe\,{\sc ii} features were $\sim 16000$ km~s$^{-1}$ and 
$\sim 14000$ km~s$^{-1}$, respectively. Then the velocities declined up to day +40  
and became nearly constant at later epochs.  
\end{abstract}
\begin{keywords}
Supernovae: general - supernovae: individual (SN2011fu)
\end{keywords}

\section{INTRODUCTION}
It is commonly recognized that core-collapse supernovae (CCSNe) represent 
the final stages of the life of massive stars ($M >$ 8\,--10 M$_{\sun}$) 
\citep{2003ApJ...591..288H, 2009MNRAS.399..559A, 2009ARA&A..47...63S}. Generally, the 
fate of  massive stars is governed by its mass, metallicity, rotation and magnetic field  
\citep{1999ApJ...522..413F, 2003ApJ...591..288H, 2005NatPh...1..147W}.
Massive stars show a wide variety in these fundamental parameters, causing diverse 
observational properties among various types of CCSNe. The presence of dominant 
hydrogen lines in the spectra of Type II SNe strongly suggests that their progenitors belong 
to massive stars which are still surrounded by significantly thick hydrogen envelope before 
the explosion \citep[see][for a review on different types of SNe]{1997ARA&A..35..309F}. 
On the contrary, Type Ib events are H-deficient but He is still present in their spectra, 
unlike to Type Ic SNe, where both H and He features are absent. After the discovery of 
SN~1987K, another class, termed as Type IIb \citep[see][]{1988AJ.....96.1941F, 1987ApJ...318..664W},
was included in the CCSN zoo, and the observational properties of these SNe 
closely resemble those of Type II SNe during the early phases, while they are more similar 
to Type Ib/c events at later epochs.

However, in a few cases, the spectral classification of Type IIb SNe is more controversial: 
for example, SN~2000H \citep{2000IAUC.7375....2B, 2002ApJ...566.1005B, 2006A&A...450..305E}; 
SN~2003bg \citep{2003IAUC.8084....4F, 2006ApJ...651.1005S}; SN 2007Y (although this event is 
classified as Type Ib/c by e.g. \citet{2007CBET..845....1M, 2009ApJ...696..713S} 
however, \citet{2010MNRAS.409.1441M} suggested that it is a Type IIb) and SN~2009mg 
\citep{2009CBET.2087....1P, 2009CBET.2093....1R, 2010CBET.2158....1S}). SNe of Type IIb  
are further divided into two subgroups: Type cIIb with compact progenitors like 
SNe 1996cb, 2001ig and 2008ax, and Type eIIb with extended progenitors, e.g SNe 1993J
and 2001gd \citep{2010ApJ...711L..40C}. 

Type IIb and Type Ib/c SNe are collectively known as ``stripped envelope'' CCSNe 
\citep{1997ApJ...483..675C} as the outer envelopes of hydrogen and/or helium of
their progenitors are partially or completely removed before the explosion. 
The possible physical mechanisms behind this process may be stellar winds 
\citep{2008A&ARv..16..209P} or interaction with a companion star in a binary system 
where mass transfer occurs due to Roche lobe overflow \citep{1992ApJ...391..246P}.
There are several studies about the discovery of the progenitors of Type IIb SNe 
but the debate about how they manage to keep only a thin layer of hydrogen, is still on
\citep{1994AJ....107..662A, 2004Natur.427..129M, 2006MNRAS.369L..32R, 
2008MNRAS.391L...5C, 2008AstBu..63..228S, 2011ApJ...742L..18A, 2011ApJ...739L..37M,
2011ApJ...741L..28V, 2012ApJ...752...78S}.

To date, approximately 
77\footnote{http:heasarc.gsfc.nasa.gov/W3Browse/star-catalog/asiagosn.html} Type IIb SNe are known,
but only a few of them have been properly monitored and well-studied. Among them, SNe 1987K
\citep{1988AJ.....96.1941F}; 1993J \citep{1993Natur.364..600S, 1994MNRAS.266L..27L, 1994AJ....107.1022R}; 
1996cb \citep{1999AJ....117..736Q}, 2003bg \citep{2009ApJ...703.1612H,2009ApJ...703.1624M}; 
2008ax \citep{2008MNRAS.389..955P, 2009ApJ...704L.118R, 2011ApJ...739...41C, 2011MNRAS.413.2140T}; 
2009mg \citep{2012MNRAS.424.1297O}; 2011ei \citep{2012arXiv1207.2152M} 
and more recently 2011dh 
\citep{2011ApJ...742L..18A, 2011A&A...535L..10M, 2011ApJ...739L..37M, 2011ApJ...741L..28V,
2012ApJ...757...31B, 2012ApJ...751..125B, 2012arXiv1209.1102H, 2012ApJ...750L..40K, 2012ApJ...752...78S, 
2012A&A...540A..93V} were remarkably well-studied.

An interesting property of the observed light curves (LCs) of a few Type IIb SNe is the initial peak and
rapid decline followed by a subsequent rise and a secondary maximum. The first peak is thought 
to be due to the break-out of the SN shock from the extended progenitor envelope 
\citep{1977ApJS...33..515F}. The properties of the shock break-out peak depend on the envelope 
mass and the density structure of the outer layers. The shock break-out phase can last from 
seconds to days. Therefore, early discovery and rapid-cadence early-time observations may help  
in understanding the properties of the outer envelope of massive stars in a better way
\citep{2011ApJ...736..159G}. 

In this paper, we present the results from photometric and spectroscopic monitoring of 
SN~2011fu starting shortly after the discovery and extending up to nebular phases.
The photometric and spectroscopic properties of this event reveal that SN~2011fu is a Type IIb
supernova. The type determination for this SN was verified with SNID \citep{2007ApJ...666.1024B},
highlighting an excellent resemblance of the object with SN~1993J.

The paper is organized as follows. The photometric and spectroscopic observations are 
presented in Section 2, where the methods for data reduction and analysis are described. 
In Section 3, we analyze the light and colour curves. In Sections 4, we
describe and model the bolometric LC. 
Section 5 deals with the spectroscopic modelling using the \texttt{SYNOW} code. In Section 6, we 
discuss the metallicites of the host galaxy of SN~2011fu along with those of other CCSNe. 
Finally, the results are summarized in Section 7.    

\begin{figure}
\centering
\includegraphics[scale = 0.4]{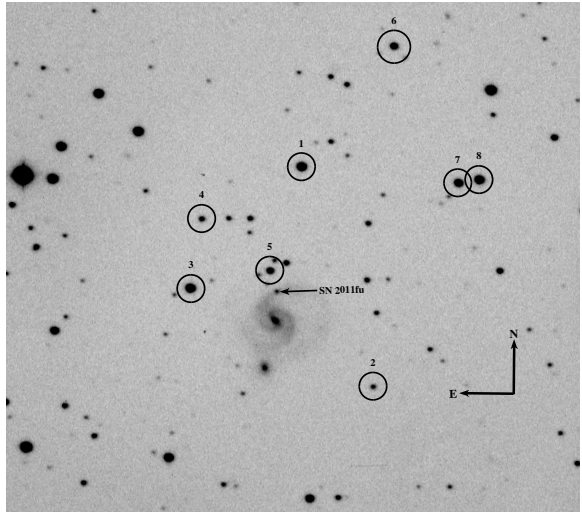}
\caption{The $V$-band image of the SN~2011fu field around the galaxy UGC 01626, 
observed on 2011 November 16 with the 1-m ST, India. The SN is marked with a black arrow. 
The reference standard stars used for calibration are marked with
numbers 1-8. On this image, north is up and east is to the left.}
\label{fig:field}
\end{figure}

\section{Observations and data analysis} \label{sec:phot}

SN~2011fu was discovered in a spiral arm of the galaxy UGC 01626 (type SAB(rs)$c$) by 
F. Ciabattari and E. Mazzoni \citep{2011CBET.2827....1C} on 2011 September 21.04 (UT) 
with a 0.5-m Newtonian telescope in the course of the Italian Supernovae Search Project. 
The brightness of the SN at the time of discovery was reported to be at mag $\sim$ 16 
(unfiltered).
It was located 2\arcsec\, west and  26\arcsec\, north of the center of the host 
galaxy, with coordinates $\alpha = 02^{\rm h} 08^{\rm m} 21\fs41$, 
$\delta =+41\degr 29\arcmin 12\farcs3$ (equinox 2000.0) \citep{2011CBET.2827....1C}.
The host galaxy has a heliocentric velocity and redshift of $5543 \pm 11$ km~s$^{-1}$ 
and $z$ = $0.018489 \pm 0.000037$\footnote{HyperLEDA - http://leda.univ-lyon1.fr}, 
respectively. The first spectrum of~SN2011fu was obtained on 2011 September 23.84 UT with
the Ekar-Copernico 1.82-m telescope (range 360-810 nm; resolution 2.2 nm) by
\citet{2011CBET.2827....2T}, showing a blue continuum with superimposed 
weak H and He\,{\sc i} 587.6-nm features, which led to the classification as a young 
Type II SN.
 
\subsection {Optical Photometry} \label{sec:phot.d}

The prompt photometric follow-up of SN 2011fu started shortly after the discovery and 
continued using three ground-based telescopes in India. The majority of the 
observations were made using the 2-m {\it Himalayan Chandra Telescope} (HCT) of the Indian Astronomical 
Observatory, Hanle and the 1-m {\it Sampurnanand Telescope} (ST) at the 
Aryabhatta Research Institute of observational sciencES (ARIES), Nainital, India.
All observations were performed in Bessell \emph{UBVRI} bands.

\begin{table*}
  \caption{Identification number (ID), coordinates ($\alpha, \delta$) and
calibrated magnitudes of standard stars in the field of SN 2011fu.}
 \label{tab:photstd}
 \begin{tabular}{cccccccc}
 \hline
 \hline
 Star& $\alpha_{\rm J2000}$ & $\delta_{\rm J2000}$  & $U$ & $B$ & $V$ &
$R$ & $I$ \\
       ID& (h m s)              &
(\degr\,\arcmin\,\arcsec)&(mag)&(mag)&(mag)&(mag)&(mag) \\
     \hline
 1&02 08 19.66&+41 30 53.4&15.59$\pm$0.02&15.47$\pm$0.02&14.80$\pm$0.01&14.38$\pm$0.02&14.03$\pm$0.02\\
 2&02 08 14.25&+41 27 51.8&18.39$\pm$0.09&18.17$\pm$0.02&17.59$\pm$0.02&17.16$\pm$0.02&16.83$\pm$0.03\\
 3&02 08 27.70&+41 29 11.4&16.29$\pm$0.02&15.70$\pm$0.02&14.86$\pm$0.02&14.34$\pm$0.02&13.93$\pm$0.03\\
 4&02 08 26.92&+41 30 07.5&17.53$\pm$0.04&17.58$\pm$0.02&17.07$\pm$0.02&16.70$\pm$0.02&16.40$\pm$0.03\\
 5&02 08 21.94&+41 29 26.9&16.78$\pm$0.03&16.77$\pm$0.02&16.21$\pm$0.01&15.84$\pm$0.02&15.54$\pm$0.02\\
 6&02 08 12.72&+41 32 34.8&16.36$\pm$0.02&16.45$\pm$0.02&15.85$\pm$0.01&15.44$\pm$0.02&15.07$\pm$0.02\\
 7&02 08 08.04&+41 30 41.3&16.32$\pm$0.02&16.07$\pm$0.02&15.38$\pm$0.01&14.97$\pm$0.02&14.61$\pm$0.02\\
 8&02 08 06.54&+41 30 43.7&17.23$\pm$0.03&16.03$\pm$0.02&14.93$\pm$0.02&14.32$\pm$0.02&13.78$\pm$0.02\\

     \hline
  \end{tabular}
\end{table*}

The HCT photometric observations started on 2011 September 28 using the Himalaya Faint Object 
Spectrograph Camera (HFOSC). The central 2k $\times$ 2k region of a 2k $\times$ 4k 
SITe CCD chip was used for imaging which provided an image scale of 0.296 arcsec 
pixel$^{-1}$ across a 10 $\times$ 10 arcmin$^2$ field-of-view. 

Further photometric observations were carried out using a 2k $\times$ 2k CCD camera at the f/13 
Cassegrain focus of the 1-m ST telescope situated at ARIES, Nainital. The CCD chip has 
square pixels of $24 \times 24$\,$\umu$m, a scale of 0.38 arcsec per pixel and the 
entire chip covers a field of 13 $\times$ 13 arcmin$^2$ on the sky. The gain and readout 
noise of the CCD camera are 10 electron per ADU and 5.3 electrons, respectively.  
A finding chart showing the field of the SN~2011fu along with the local standard stars 
is presented in Fig.~\ref{fig:field}.

In addition, we also observed this SN in $V$ and $R$ bands on 2011 December 01 and 
2012 March 02 using the 1.3m telescope DFOT \footnote{DFOT uses 
2048 $\times$ 2048 ANDOR CCD camera having $13.5 \times 13.5$ $\umu$m pixels mounted 
at the f/4 Cassegrain focus of the telescope. With 0.54 arcsec per pixel plate scale, 
the entire chip covers a 18 $\times$ 18 arcmin$^2$ field-of-view on the sky. The CCD 
can be read out with 31, 62, 500 and 1000 kHz speed, with system RMS noise of 2.5, 4.1, 
6.5, 7 electrons and gain of 0.7, 1.4, 2, 2 electron/ADU respectively. We selected the 
500 kHz readout frequency during our observations.} 
\citep{Sagar2011Csi...101...8.25, 2012ASInC...1..113S}, recently installed at Devasthal, 
Naintial. 

To improve the signal-to-noise ratio (S/N), all the photometric observations were carried
out with 2$\times$2 binning.  
Along with science frames several bias and twilight flat frames were also collected.
Alignment and determination of mean FWHM on all science frames were performed 
after the usual bias subtraction, flat fielding and cosmic-ray removal. The standard 
tasks available in {\sc iraf\,}\footnote{{\sc iraf} stands for Image Reduction and 
Analysis Facility distributed by the National Optical Astronomy Observatories which
is operated by the Association of Universities for research in Astronomy,
Inc. under co-operative agreement with the National Science Foundation.}
and {\sc daophot}\footnote{ {\sc daophot} stands for Dominion Astrophysical 
Observatory Photometry.} \citep{1987PASP...99..191S, 1992ASPC...25..297S} were used 
for pre-processing and photometry.

The pre-processing steps for images taken with all three telescopes were performed in a similar fashion. 
The stellar FWHM on the $V$-band frames typically varied from 2\arcsec\, to 4\arcsec, with a
median value of around 2\farcs5. We also co-added individual frames, wherever necessary, before
computing the final photometry. 

For photometric calibration, we observed the standard field PG0231 \citep{2009AJ....137.4186L} 
in \emph{UBVRI} bands with the 1-m ST on 2011 December 17 under good photometric conditions
(transparent sky, seeing FWHM in $V \sim 2\arcsec$). The profile fitting technique was
applied for the photometry of SN~2011fu and Landolt field and then instrumental magnitudes
were converted into standard system following least-square linear regression procedures
outlined in \citet{1992ASPC...25..297S}. 
Atmospheric extinction values (0.57, 0.28, 0.17, 0.11 and 0.07 mag per unit airmass 
for $U$, $B$, $V$, $R$ and $I$ bands, 
respectively) for the site were adopted from \citet{2000BASI...28..675K}. The chosen Landolt stars
for calibration had the brightness range of $12.77 \leq V \leq 16.11$ mag and colour range
of $-0.33 \leq B-V \leq 1.45 $ mag.  Using these stars, transformation to the standard system was 
derived by applying the following zero-points and colour coefficients:

 \[ u -U =(7.27\pm0.01) + (-0.08\pm0.02)(U-B) \]
 \[ b -B =(4.90\pm0.004) + (-0.04\pm0.01)(B-V) \]
 \[ v -V =(4.34\pm0.01) + (-0.04\pm0.01)(B-V) \]
 \[ r -R =(4.19\pm0.01) + (-0.04\pm0.01)(V-R) \]
 \[ i -I =(4.60\pm0.02) + (0.04\pm0.02)(V-I) \]

Here \emph{U}, \emph{B}, \emph{V}, \emph{R}, \emph{I} are the catalogue magnitudes and $u$, $b$, $v$,
$r$, $i$ are the corresponding instrumental magnitudes. 
Table~\ref{tab:photstd} lists the coordinates and magnitudes of the eight local
secondary standard stars in the SN field.

To estimate the possible contribution from the host galaxy to the measured supernova 
fluxes, we used the {\it ISIS}\footnote{http://www2.iap.fr/users/alard/package.html} 
image subtraction package. We acquired deep images (having total exposure times 
of more than 20 minutes) in $BVRI$ bands with the HCT telescope on 25 August 2012 
under good sky conditions. As the supernova was not detected in anyone of these frames, 
we used them as template frames for image subtraction. We found minor differences, 
not exceeding 0.1 mag, between the SN magnitudes with and without applying the image 
subtraction for the data at later epochs i.e. 70 days after the first observation. 
The final results of our SN photometry (without applying image subtraction corrections) 
along with robustly determined PSF errors, are presented in Table~\ref{tab:photsn}.

\begin{table*}
  \caption{Photometric observational log of SN 2011fu}
  \label{tab:photsn}
\begin{tabular}{cccccccc}
     \hline
     \hline
     JD& Phase$^{a}$ & $U$ & $B$ & $V$ & $R$ & $I$ &Telescope \\
       & (Days)&(mag)&(mag)&(mag)&(mag)&(mag) \\
     \hline
2455833.23 & +10.73  & 17.36$\pm$ 0.03& 17.68 $\pm$ 0.02&  17.35$\pm$ 0.01 & 16.99 $\pm$ 0.02 & 16.74 $\pm$ 0.02 & HCT \\    
2455834.49 & +11.99  & 17.66$\pm$ 0.03& 17.87 $\pm$ 0.02&  17.48$\pm$ 0.01 & 17.11 $\pm$ 0.02 & 16.86 $\pm$ 0.02 & HCT \\    
2455836.15 & +13.65  & --             & 17.92 $\pm$ 0.03&  17.46$\pm$ 0.01 & 17.08 $\pm$ 0.02 & 16.88 $\pm$ 0.02 & HCT \\    
2455837.26 & +14.76  & 17.73$\pm$ 0.03& 17.89 $\pm$ 0.02&  17.42$\pm$ 0.01 & 17.02 $\pm$ 0.02 & 16.83 $\pm$ 0.02 & HCT \\    
2455841.23 & +18.73  & 17.65$\pm$ 0.03& 17.65 $\pm$ 0.02&  17.15$\pm$ 0.01 & 16.77 $\pm$ 0.02 & 16.59 $\pm$ 0.02 & ST  \\    
2455842.30 & +19.80  & 17.69$\pm$ 0.05& 17.63 $\pm$ 0.03&  17.07$\pm$ 0.01 & 16.70 $\pm$ 0.02 & 16.56 $\pm$ 0.03 & ST  \\    
2455843.27 & +20.77  & 17.62$\pm$ 0.09& 17.51 $\pm$ 0.05&  17.05$\pm$ 0.02 & 16.67 $\pm$ 0.02 & 16.57 $\pm$ 0.04 & ST  \\    
2455844.21 & +21.71  & 17.48$\pm$ 0.06& 17.49 $\pm$ 0.03&  17.01$\pm$ 0.02 & 16.62 $\pm$ 0.02 & 16.48 $\pm$ 0.03 & ST  \\    
2455845.44 & +22.94  & 17.43$\pm$ 0.03& 17.48 $\pm$ 0.03&  16.95$\pm$ 0.02 & 16.59 $\pm$ 0.02 & 16.45 $\pm$ 0.03 & ST  \\    
2455845.43 & +22.93  & --             & 17.43 $\pm$ 0.02&  16.93$\pm$ 0.01 & 16.54 $\pm$ 0.02 & 16.43 $\pm$ 0.02 & HCT \\    
2455846.44 & +23.94  & 17.51$\pm$ 0.03& 17.47 $\pm$ 0.03&  16.92$\pm$ 0.02 & 16.57 $\pm$ 0.02 & 16.43 $\pm$ 0.03 & ST  \\    
2455846.43 & +23.93  & --             & 17.45 $\pm$ 0.02&  16.92$\pm$ 0.01 & 16.54 $\pm$ 0.02 & 16.45 $\pm$ 0.02 & HCT \\    
2455849.41 & +26.91  & 17.68$\pm$ 0.03& 17.58 $\pm$ 0.03&  16.95$\pm$ 0.01 & 16.53 $\pm$ 0.02 & 16.38 $\pm$ 0.02 & HCT \\    
2455850.40 & +27.89  & 17.76$\pm$ 0.04& 17.65 $\pm$ 0.03&  16.96$\pm$ 0.01 & 16.57 $\pm$ 0.02 & 16.43 $\pm$ 0.03 & HCT \\    
2455851.31 & +28.81  & 18.21$\pm$ 0.13& 17.82 $\pm$ 0.04&  17.06$\pm$ 0.03 & 16.56 $\pm$ 0.02 & 16.44 $\pm$ 0.03 & ST  \\    
2455857.30 & +34.80  & 18.99$\pm$ 0.06& 18.52 $\pm$ 0.03&  17.44$\pm$ 0.01 & 16.80 $\pm$ 0.02 & 16.58 $\pm$ 0.03 & ST  \\    
2455858.32 & +35.82  & 19.00$\pm$ 0.09& 18.60 $\pm$ 0.03&  17.47$\pm$ 0.01 & 16.84 $\pm$ 0.02 & 16.61 $\pm$ 0.03 & ST  \\    
2455859.18 & +36.68  & --             & 18.74 $\pm$ 0.03&  17.51$\pm$ 0.01 & 16.92 $\pm$ 0.02 & 16.69 $\pm$ 0.02 & HCT \\    
2455860.30 & +37.80  & 19.10$\pm$ 0.08& 18.73 $\pm$ 0.03&  17.60$\pm$ 0.01 & 16.93 $\pm$ 0.02 & 16.68 $\pm$ 0.02 & ST  \\    
2455862.30 & +39.80  & 19.24$\pm$ 0.15& 18.86 $\pm$ 0.03&  17.69$\pm$ 0.02 & 16.97 $\pm$ 0.02 & 16.71 $\pm$ 0.03 & ST  \\    
2455864.40 & +41.90  & 19.55$\pm$ 0.08& 19.05 $\pm$ 0.03&  17.77$\pm$ 0.01 & 17.09 $\pm$ 0.02 & 16.84 $\pm$ 0.02 & HCT \\    
2455865.35 & +42.85  & --             & 19.12 $\pm$ 0.02&  17.81$\pm$ 0.01 & 17.11 $\pm$ 0.02 & 16.85 $\pm$ 0.02 & HCT \\    
2455866.23 & +43.73  & --             & 18.97 $\pm$ 0.05&  17.88$\pm$ 0.02 & 17.14 $\pm$ 0.02 & 16.83 $\pm$ 0.03 & ST  \\    
2455866.26 & +43.76  & --             & 19.10 $\pm$ 0.02&  17.86$\pm$ 0.02 & 17.17 $\pm$ 0.02 & 16.88 $\pm$ 0.02 & HCT \\    
2455875.22 & +52.72  & --             & --              &  18.06$\pm$ 0.06 & 17.33 $\pm$ 0.04 & 16.95 $\pm$ 0.05 & ST  \\    
2455879.28 & +56.78  & --             & 19.30 $\pm$ 0.10&  18.17$\pm$ 0.03 & 17.55 $\pm$ 0.03 & 17.18 $\pm$ 0.03 & ST  \\    
2455881.26 & +58.76  & --             & --              &  18.19$\pm$ 0.02 & 17.50 $\pm$ 0.02 & 17.18 $\pm$ 0.03 & HCT \\    
2455882.33 & +59.83  & --             & 19.38 $\pm$ 0.07&  18.19$\pm$ 0.03 & 17.56 $\pm$ 0.02 & 17.20 $\pm$ 0.03 & ST  \\    
2455884.27 & +61.78  & 19.94$\pm$ 0.05& 19.49 $\pm$ 0.03&  18.28$\pm$ 0.01 & 17.62 $\pm$ 0.01 & 17.30 $\pm$ 0.02 & HCT \\    
2455894.23 & +71.73  & 19.67$\pm$ 0.20& 19.37 $\pm$ 0.05&  18.41$\pm$ 0.03 & 17.76 $\pm$ 0.02 & 17.42 $\pm$ 0.03 & ST  \\    
2455896.28 & +73.78  & 19.75$\pm$ 0.07& 19.49 $\pm$ 0.03&  18.43$\pm$ 0.01 & 17.87 $\pm$ 0.02 & 17.57 $\pm$ 0.03 & HCT \\    
2455897.08 & +74.58  & --             & --              &  18.39$\pm$ 0.03 & 17.79 $\pm$ 0.03 & --               & DFOT\\    
2455898.30 & +75.80  & --             & 19.28 $\pm$ 0.05&  18.41$\pm$ 0.02 & 17.79 $\pm$ 0.02 & 17.45 $\pm$ 0.03 & ST  \\    
2455900.17 & +77.66  & --             & 19.47 $\pm$ 0.08&  18.50$\pm$ 0.03 & 17.83 $\pm$ 0.03 & 17.53 $\pm$ 0.03 & ST  \\    
2455901.24 & +78.74  & --             & 19.36 $\pm$ 0.06&  18.48$\pm$ 0.04 & 17.88 $\pm$ 0.03 & 17.39 $\pm$ 0.03 & ST  \\    
2455904.28 & +81.78  & --             & 19.31 $\pm$ 0.17&  --              & 17.97 $\pm$ 0.05 & 17.72 $\pm$ 0.04 & HCT \\    
2455909.18 & +86.68  & --             & 19.52 $\pm$ 0.07&  18.60$\pm$ 0.03 & 17.96 $\pm$ 0.03 & 17.63 $\pm$ 0.03 & ST  \\    
2455912.28 & +89.78  & --             & 19.45 $\pm$ 0.05&  18.64$\pm$ 0.03 & 18.06 $\pm$ 0.03 & 17.67 $\pm$ 0.04 & ST  \\    
2455913.24 & +90.73  & 19.56$\pm$ 0.12& 19.47 $\pm$ 0.05&  18.59$\pm$ 0.03 & 18.09 $\pm$ 0.03 & 17.70 $\pm$ 0.03 & ST  \\    
2455918.18 & +95.68  & --             & 19.55 $\pm$ 0.03&  --              & 18.21 $\pm$ 0.03 & 17.82 $\pm$ 0.03 & HCT \\    
2455919.11 & +96.61  & --             & 19.58 $\pm$ 0.04&  18.75$\pm$ 0.01 & 18.22 $\pm$ 0.02 & 17.87 $\pm$ 0.02 & HCT \\    
2455922.17 & +99.67  & --             & 19.48 $\pm$ 0.06&  18.74$\pm$ 0.03 & 18.24 $\pm$ 0.03 & 17.81 $\pm$ 0.04 & ST  \\    
2455924.10 & +101.60 & --             & 19.65 $\pm$ 0.03&  18.86$\pm$ 0.01 & 18.34 $\pm$ 0.02 & 18.02 $\pm$ 0.02 & HCT \\    
2455929.17 & +106.67 & --             & 19.73 $\pm$ 0.09&  18.92$\pm$ 0.04 & 18.36 $\pm$ 0.05 & 18.00 $\pm$ 0.04 & ST  \\    
2455930.27 & +107.76 & --             & --              &  19.04$\pm$ 0.11 & 18.47 $\pm$ 0.07 & 17.99 $\pm$ 0.12 & ST  \\    
2455930.27 & +108.73 & --             & --              &  18.89$\pm$ 0.15 & 18.29 $\pm$ 0.10 & 17.93 $\pm$ 0.10 & ST  \\    
2455932.15 & +109.65 & --             & 19.80       0.16&  19.00$\pm$ 0.05 & --               & 18.11 $\pm$ 0.03 & HCT \\    
2455936.25 & +113.75 & --             & --              &  --              & --               & 18.27 $\pm$ 0.06 & HCT \\    
2455937.16 & +114.66 & --             & 19.69 $\pm$ 0.09&  19.09$\pm$ 0.04 & 18.58 $\pm$ 0.03 & 18.13 $\pm$ 0.04 & HCT \\    
2455938.08 & +115.58 & --             & 19.70 $\pm$ 0.04&  19.11$\pm$ 0.02 & 18.65 $\pm$ 0.03 & 18.35 $\pm$ 0.03 & HCT \\    
2455939.23 & +116.73 & --             & --              &  19.00$\pm$ 0.07 & --               & 18.11 $\pm$ 0.05 & ST  \\    
2455947.06 & +124.56 & --             & --              &  19.21$\pm$ 0.02 & 18.74 $\pm$ 0.02 & 18.46 $\pm$ 0.04 & HCT \\    
2455947.16 & +124.66 & --             & --              &  --              & 18.55 $\pm$ 0.04 & 18.16 $\pm$ 0.07 & ST  \\    
2455953.08 & +130.58 & --             & 19.64 $\pm$ 0.08&  --              & 18.68 $\pm$ 0.03 & 18.38 $\pm$ 0.06 & ST  \\    
2455954.16 & +131.66 & --             & --              &  19.33$\pm$ 0.02 & 18.87 $\pm$ 0.02 & 18.49 $\pm$ 0.03 & HCT \\    
2455963.14 & +140.64 & --             & --              &  19.23$\pm$ 0.07 & --               & 18.54 $\pm$ 0.10 & ST  \\    
2455967.11 & +144.61 & --             & --              &  19.18$\pm$ 0.07 & 18.87 $\pm$ 0.05 & 18.48 $\pm$ 0.08 & ST  \\    
2455969.09 & +146.59 & --             & --              &  --              & 18.91 $\pm$ 0.05 & --               & ST  \\    
2455976.07 & +153.56 & --             & --              &  19.40$\pm$ 0.03 & 18.88 $\pm$ 0.03 & 18.58 $\pm$ 0.07 & ST  \\    
2455979.12 & +156.62 & --             & --              &  --              & 18.92 $\pm$ 0.05 & 18.51 $\pm$ 0.09 & ST  \\    
2455989.08 & +166.58 & --             & --              &  19.59$\pm$ 0.09 & 19.24 $\pm$ 0.10 & 18.87 $\pm$ 0.08 & DFOT, ST \\
2455998.08 & +175.58 & --             & --              &  --              & 19.07 $\pm$ 0.05 & 18.89 $\pm$ 0.10 & ST  \\    
2456159.34 & +336.84 & --            & \textgreater 22.5 & \textgreater 22 & \textgreater 21.5 & \textgreater 21 & HCT \\    
  \hline
  \end{tabular}  \\
 $^{a}$ with reference to the explosion epoch JD 2455822.5\\

 HCT : 2-m Himalayan Chandra Telescope, IAO, Hanle; DFOT : 1.3-m Devasthal
 Fast Optical Telescope, ARIES, India;
 ST : 1-m Sampurnanand Telescope, ARIES, India \\
\end{table*}

\begin{figure*}
\centering
\includegraphics[scale = 0.8]{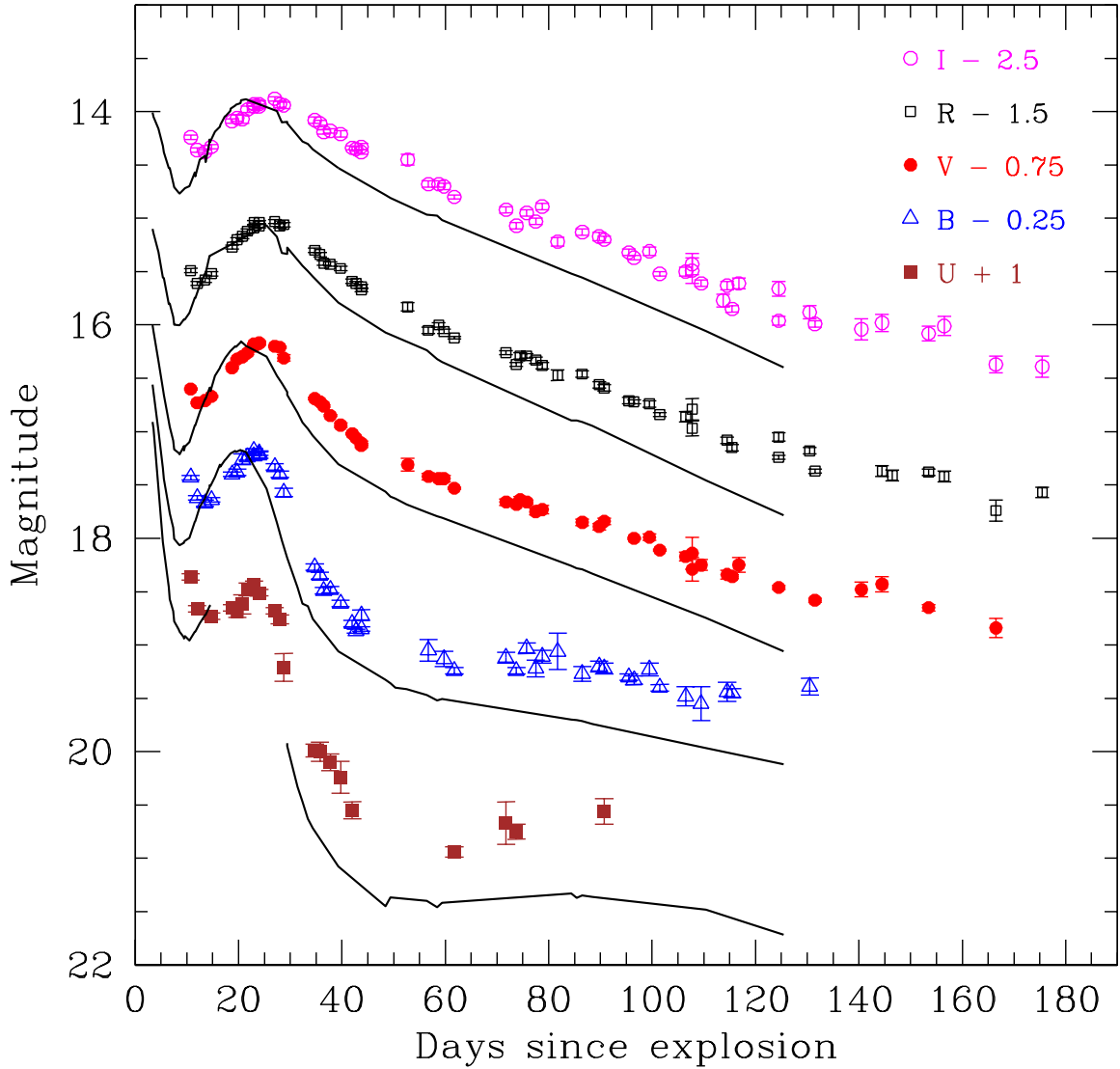}
\caption{Observed \emph{UBVRI} light curves of SN~2011fu. For clarity, the
light curves in different bands have been shifted vertically by the values indicated
in the legend. Black solid lines represent the light curves of SN~1993J over-plotted
with appropriate shifts. The explosion date of SN~2011fu was taken to be 2011 September 
$18 \pm 2$, as described in subsection 3.1.}

\label{fig:snid}
\end{figure*}

\subsection{Spectroscopy} 
Spectroscopic observations of SN 2011fu were obtained at 8 epochs, between 
2011 September 28 (JD 2455833.27) and December 22 (JD 2455918.11). A journal of these
observations is given in Table~\ref{tab_spec}. The SN spectra were taken with the HFOSC
instrument mounted at the 2-m Himalayan Chandra Telescope. All spectra were obtained 
using grisms Gr$\#$7 (wavelength range 3500 - 7800 \AA) and Gr$\#$8 (wavelength range 
5200 - 9200 \AA). FeAr and FeNe arc lamp spectra were applied for wavelength calibration. 
Spectrophotometric standard were also observed with a broader slit to correct for 
the instrumental response and flux calibration. 

\begin{table*}
\centering
\caption{Log of spectroscopic observations of SN 2011fu.}
\begin{tabular}{ccccc} \hline \hline
    Date    &     J.D.   & Phase$^{a}$& Range                & Resolution  \\ 
            &            & (Days)     & (\AA)                & (\AA)       \\ \hline
 2011-09-28 & 2455833.28 & +10.78     & 3500-7800; 5200-9250 & 7   \\ 
 2011-09-29 & 2455834.41 & +11.92     & 3500-7800; 5200-9250 & 7   \\ 
 2011-10-01 & 2455836.23 & +13.73     & 3500-7800; 5200-9250 & 7   \\ 
 2011-10-14 & 2455849.43 & +26.93     & 3500-7800; 5200-9250 & 7   \\ 
 2011-10-29 & 2455864.41 & +41.91     & 3500-7800; 5200-9250 & 7   \\ 
 2011-10-31 & 2455866.36 & +43.86     & 3500-7800; 5200-9250 & 7   \\ 
 2011-11-23 & 2455889.15 & +66.65     & 3500-7800; 5200-9250 & 7   \\ 
 2011-12-22 & 2455918.11 & +95.61     & 3500-7800            & 7   \\ 
\hline
\end{tabular} \\
$^{a}$ with reference to the explosion epoch JD 2455822.5\\
\label{tab_spec}
\end{table*}

The reduction of the spectroscopic data were carried out in a standard manner using various
tasks available within {\sc iraf}. First, all images were bias-subtracted and flat fielded.
Then, one dimensional spectra were extracted from the two-dimensional cleaned images
using the optimal extraction algorithm \citep{1986PASP...98..609H}. The wavelength 
calibration was computed using the arc spectra mentioned above. The accuracy of the 
wavelength calibration was checked using the night sky emission lines and small shifts 
were applied to the observed spectra whenever required. The instrumental response 
curves were determined using the spectrophotometric standards observed on the same 
night as the SN, and the SN spectra were calibrated to a relative flux scale. When the 
spectrophotometric standards could not be observed, the response curve based on 
observations in a night close in time to the SN observation was adopted.
The flux calibrated spectra in the two regions were combined to a weighted mean 
to obtain the final spectrum on a relative flux scale.
 
Finally, the spectra were brought to an absolute flux scale using zero points determined
from the calibrated, broad-band \emph{UBVRI} magnitudes. The SN spectra were also
corrected for the redshift of the host galaxy (z = 0.018), and de-reddened 
assuming a total reddening of $E(B-V)$ = 0.22 mag (see Sub-section 3.3). 
The telluric lines have not been removed from the spectra.

\begin{table*}
\centering
\caption{Epochs of the LC valley ($t_v$) and the secondary peak ($t_p$)
in days after explosion, and their respective apparent magnitudes for SN 2011fu
and SN 1993J.
\label{tab:peaks}}
\begin{tabular}{cccccc}
\hline \hline
SN      &Band    & LC valley      & Apparent magnitude& LC peak      & Apparent magnitude \\
        &        & $t_v$ (days)   & at $t_v$          & $t_p$ (days) & at $t_p$           \\ \hline
        &  U     & 15.51$\pm$4.34 & 17.67$\pm$0.42 & 22.93$\pm$3.64 &17.43$\pm$3.34 \\
        &  B     & 13.75$\pm$1.47 & 17.93$\pm$0.80 & 23.29$\pm$2.89 &17.51$\pm$0.83 \\
2011fu  &  V     & 12.87$\pm$1.69 & 17.45$\pm$1.32 & 24.96$\pm$2.01 &16.95$\pm$0.42 \\
        &  R     & 12.95$\pm$1.81 & 17.01$\pm$1.20 & 26.40$\pm$2.90 &16.50$\pm$0.11 \\
        &  I     & 13.50$\pm$1.89 & 16.86$\pm$0.67 & 26.64$\pm$2.80 &16.41$\pm$0.32 \\ \hline
        &  U     & 10.33$\pm$1.52 & 11.94$\pm$0.76 &         --     &       --      \\
        &  B     & 8.82 $\pm$3.36 & 12.27$\pm$1.74 & 19.92$\pm$0.70 &11.40$\pm$0.17 \\
1993J   &  V     & 8.96 $\pm$1.41 & 11.89$\pm$1.14 & 21.67$\pm$0.66 &10.87$\pm$0.12 \\
        &  R     & 8.81 $\pm$1.06 & 11.47$\pm$0.61 & 22.53$\pm$3.24 &10.52$\pm$0.37 \\
        &  I     & 9.17 $\pm$1.58 & 11.25$\pm$0.93 & 23.06$\pm$1.91 &10.39$\pm$0.21 \\
\hline
\end{tabular}
\end{table*}

\section{Light curves of SN~2011fu}

In this section, we present the multi-band light curves of SN~2011fu and their
comparison with SN~1993J light curves and their temporal properties.
A brief discussion about the explosion epoch of SN~2011fu is presented in the
following sub-section.

\subsection{Explosion epoch of SN~2011fu}

The detection of very early time light curve features of SN~2011fu, similar to 
that seen in case of SN~1993J, indicates a very young age at the time of discovery. 
Very sharp rise followed by a relatively fast decline are explained as 
the detection of cooling phase and depends mainly on the $^{56}$Ni mixing and the progenitor 
radius, as shown by hydrodynamical models of H-stripped CCSNe  
\citep{1994ApJ...420..341S, 1994ApJ...429..300W, 1998ApJ...496..454B, 2012ApJ...757...31B}.
For example, in the case of SN 2011dh, for progenitor radius of $\textless$ 300 $R_\odot$, the cooling 
phase ends at $\sim$ 5 days after the explosion \citep[see Figure-10 of][]{2012ApJ...757...31B}.

In the literature, the first detection of SN~2011fu has been reported to be 2011
September 20.708 (Z. Jin and X. Gao, Mt. Nanshan, China). However, according to 
\citet{2011CBET.2827....1C}, this object was not visible on 2011 August 10 at SN location, 
putting a stringent limit to the explosion date. We collected following pieces of evidences to 
put a constrain on the explosion date of~SN 2011fu.

\begin{enumerate}
\item For CCSNe of Type Ib and IIb, the explosion dates have been estimated to be $\sim$ 20
days prior to the $V$-band maxima \citep{2006AJ....131.2233R, 2011ApJ...741...97D} 
\citep[see also][]{2012arXiv1207.2152M}.
\item Type IIb SNe also exhibit bluer $B-V$ colour $\sim$ 40 days after the explosion 
\citep{2008MNRAS.389..955P}, giving an indication about the explosion epoch.
\item The SNID \citep{2007ApJ...666.1024B} fitting on initial four spectra of SN~2011fu 
indicates that explosion of this event would have occurred around 2011 September 20. 
However, SNID fit for the later three epochs of spectra (after V band maximum) gives rise 
to 2011 September 17 as the explosion date.
\item In some of the well studied type IIb SNe, the explosion epoch is better constrained
(e.g SN~1993J, SN~2008ax and SN~2011dh) and their early light curve features indicate
that the adiabatic cooling phase may be observable for several days after the explosion
and this duration depends upon the volume of the photospheric shell \citep{2009ApJ...704L.118R}, 
as determined in case of SN 1993J 
\citep{1993ApJ...417L..71W, 1994MNRAS.266L..27L, 1995A&AS..110..513B}, 
SN~2008ax \citep{2009ApJ...704L.118R} and SN~2011dh \citep{2011ApJ...742L..18A}.
\end{enumerate}

Based on above evidences, we have adopted 2011 September $18 \pm 2$ as the explosion epoch for
SN~2011fu and it will be used for the further discussions in this article.

\subsection{Light curve analysis}

In Fig.~\ref{fig:snid}, we plot the calibrated \emph{UBVRI} light curves of SN~2011fu.
The LCs span $\sim$ 175 days after the explosion. It is clear from 
Fig.~\ref{fig:snid} that the photometric observations of this supernova started shortly 
after explosion, showing the early declining phase in all bands, which is possibly 
related to the cooling tail after the shock break-out from an extended progenitor envelope
\citep{1992ApJ...394..599C, 2007ApJ...667..351W, 2008ApJ...683L.135C, 2010ApJ...725..904N}. 
The LCs of SN~2011fu are strikingly similar to those of SN~1993J, both in the 
initial and the following phases, exhibiting valley-like structures followed by rising 
peaks in all bands. At late epochs the LCs are monotonically decreasing in all bands, 
as expected for expanding, cooling ejecta heated by only the radioactive decay 
of $^{56}$Ni and $^{56}$Co. 

Beside SNe 1993J and 2011dh, SN~2011fu is the third known case among IIb SNe  
to date where all the initial decline phase, the rise of the broader secondary peak and
the final decline have been observed (although \citet{2010AAS...21534203R} reported 
similar observations for SN~2008ax). In the followings, we refer the first minimum of the 
LC (when the initial decline stops and the rise to the secondary maximum starts) as 
the ``valley".   

To determine the epochs of the valleys ($t_v$, in days), the subsequent peaks 
($t_p$, in days) and their corresponding brightness values, we fitted a third-order 
polynomial using a $\chi^2$ minimization technique to the LCs of both SN~2011fu and 
SN~1993J. The errors in the fitting procedure were estimated by the error propagation method. 
We have taken 1993 March 27.5 as explosion date for SN~1993J \citep{1993ApJ...417L..71W}.
The derived values of $t_v$, $t_p$ and corresponding brightness values for both SNe are 
listed in Table~\ref{tab:peaks}. 

The values of $t_v$ and $t_p$ for both the SNe are similar within the errors in all the bands.
However, for both SNe, the light curves peak earlier in blue bands than in the 
red bands (see Table~\ref{tab:peaks}) which is a common feature seen in CCSNe.
By applying linear regression method, the decline and rising rates (in mag day$^{-1}$) were also
estimated for three phases, i.e. the pre-valley ($\alpha_1$), valley-to-peak ($\alpha_2$) and 
after-peak phases ($\alpha_3$). The results of the fitting are shown in Table~\ref{tab:rates}. 
These values suggest that for SN~2011fu the pre-valley decay rates ($\alpha_1$) 
are steeper (i.e. the decay is faster) at shorter wavelengths. This is also true for SN~1993J, 
where the decay rates ($\alpha_1$) were even steeper. Thus, the initial LC decay of SN~1993J 
was steeper than that of SN~2011fu during this early 
phase \citep[see also][]{1995A&AS..110..513B}.
Between valley to peak phase ($\alpha_2$), the LC of SN~2011fu evolved with a similar rate 
in all the bands, but slower than that seen in case of SN~1993J. During the post-peak phase,
the LCs gradually became flatter at longer wavelengths (see $\alpha$$_3$ values in 
Table~\ref{tab:rates}). 
This trend has also been observed for SN~1993J and other Type IIb SNe. The $B$-band LC of 
SN~2011fu between 50 and 100 days after explosion might even show a plateau, similar to 
SNe~1993J \citep[see Fig. 3 of][]{1994MNRAS.266L..27L} and 1996cb \citep[see Fig. 2 and 
the discussions of][]{1999AJ....117..736Q}. The plateau-like behaviour of the $U$-band 
LC of SN~2011fu event is more prominent than the $U$-band LC of SN~1993J. 

We also determined the $\Delta m_{15}$ parameter for the $V$-band LCs of both SNe, 
$\Delta m_{15}$ is defined as the decline in magnitude after 15 days post-maximum.  
We got $\Delta m_{15}$(V) = 0.75 mag for SN~2011fu which is slightly lower than that 
for SN~1993J ($\Delta m_{15}$(V) = 0.9 mag). Both of these values are consistent with the mean
$\Delta m_{15}$(V) $\sim$ 0.8 $\pm0.1$ mag for Type Ib/c SNe \citep{2011ApJ...741...97D}.

\begin{table*}
\centering
\caption{Magnitude decay rate (in mag day$^{-1}$) before valley ($\alpha$$_1$), rising rate
between valley to peak ($\alpha$$_2$) and decay rate after the peak ($\alpha$$_3$) in 
SN 2011fu and SN 1993J.
\label{tab:rates}}
\begin{tabular}{ccccc}
\hline \hline
   SN     & Band   & Decay rate       &  Rising rate between & Decay rate     \\
          &        & before valley    &  valley to peak      & after peak     \\
          &        & ($\alpha$$_1$)   &   ($\alpha$$_2$)     & ($\alpha$$_3$)  \\
\hline
          &  U     & 0.24 $\pm$ 0.05  &$-$0.04  $\pm$ 0.01    &  0.13   $\pm$0.01\\
          &  B     & 0.15 $\pm$ 0.02  &$-$0.05  $\pm$ 0.01    &  0.10   $\pm$0.01\\
  2011fu  &  V     & 0.11 $\pm$ 0.03  &$-$0.06  $\pm$ 0.01    &  0.05   $\pm$0.01\\
          &  R     & 0.09 $\pm$ 0.02  &$-$0.05  $\pm$ 0.01    &  0.04   $\pm$0.01\\
          &  I     & 0.09 $\pm$ 0.02  &$-$0.03  $\pm$ 0.01    &  0.02   $\pm$0.01\\
\hline
          &  U     & 0.38 $\pm$ 0.03  &          --           &          --      \\
          &  B     & 0.24 $\pm$ 0.02  &$-$0.08  $\pm$ 0.01    &  0.11   $\pm$0.01\\
  1993J   &  V     & 0.24 $\pm$ 0.01  &$-$0.10  $\pm$ 0.01    &  0.06   $\pm$0.01\\
          &  R     & 0.20 $\pm$ 0.01  &          --           &          --      \\
          &  I     & 0.16 $\pm$ 0.01  &$-$0.09  $\pm$ 0.01    &  0.05   $\pm$0.01\\

\hline
\end{tabular}
\end{table*}

\begin{figure}
\centering
\includegraphics[scale = 0.4]{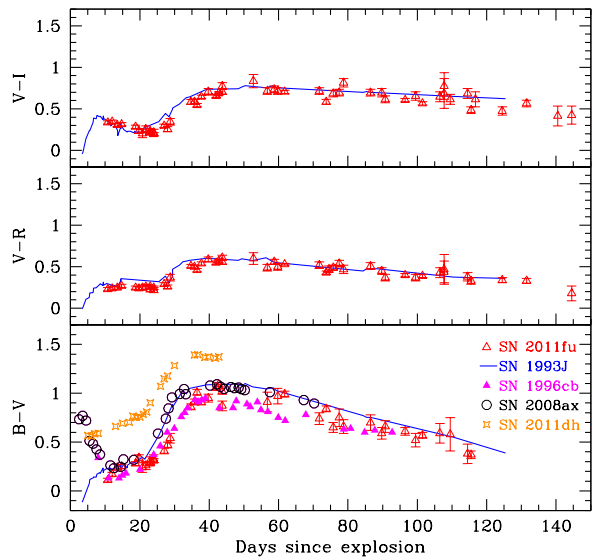}
\caption{Colour curves of SN~2011fu and other Type IIb SNe.
Bottom panel: $B-V$ colour evolution for SNe 2011fu, 2011dh, 2008ax, 1996cb (symbols)
and 1993J (blue line). Middle panel: $V-R$ colour for SN 2011fu and SN 1993J.
Top panel: the same as below but for the $V-I$ colour.}
\label{fig:color}
\end{figure}

\subsection{Colour evolution and reddening towards SN~2011fu}

In Fig.~\ref{fig:color}, we compare the evolution of the optical colour indices of 
SN~2011fu with those of other Type IIb SNe.
While constructing the colour curves, we interpolated the measured data points
(listed in Table~\ref{tab:photsn}) wherever necessary. Before plotting the colours,
reddening corrections were applied to all the bands. $E(B-V)$ = 0.068 mag was adopted as 
the reddening due to Milky Way interstellar matter (ISM) in the direction of SN~2011fu 
\citep{1998ApJ...500..525S}. The empirical correlation given by \citet{1997A&A...318..269M} 
was used to estimate the SN host galaxy extinction based on the measured Na\,{\sc i} D 
lines. For this purpose we calculated the weighted equivalent width (EW) of the un-resolved 
Na\,{\sc i} D absorption feature in the three spectra (taken on 2011 Oct 01, 14 and 31, see the log
in Table 3), 
resulted in EW (Na\,{\sc i} D) $\sim 0.35$ $\pm 0.29 $ \AA. This corresponds to 
$E(B-V)$  $\sim$ 0.15 $\pm 0.11 $ mag according to the relation given by
\citet{1997A&A...318..269M}. 
Finally, we adopted the sum of the two components, total $E(B-V)$ = 0.22 $\pm 0.11 $ mag 
as the reddening in the direction of SN~2011fu.

The bottom panel of Fig.~\ref{fig:color} shows the $B-V$ colour evolution of SN~2011fu 
along with that of SNe~1993J \citep{1994MNRAS.266L..27L}, 
1996cb \citep{1999AJ....117..736Q}, 2008ax \citep{2008MNRAS.389..955P} 
and 2011dh \citep{2012A&A...540A..93V}. It is seen in Fig.~\ref{fig:color} that the 
colour curves of SN~2011fu are similar to those of the majority of well-observed Type 
IIb SNe, except SN~2011dh which looks being redder than the others.  
 
Similar to SN 1993J, the initial $B-V$ colour of SN~2011fu increased (reddened) 
during the first 10 days (note that during the same phase SNe 2008ax and 1996cb 
showed the opposite trend). Between days +10 and +40, the $B-V$ colour continued 
to redden, then after day +40 it started to decrease and became bluer until the 
end of the our observations. This kind of colour evolution seems to be a common 
trend for Type Ib/c and IIb SNe. It may suggest that the SN ejecta became optically 
thin after 40 days. The $V-R$ (middle pannel) and $V-I$ (upper pannel) colour 
indices evolve with a similar trend as the $B-V$ colour.

\subsection{Comparison of absolute magnitudes}

The distribution of absolute magnitudes of CCSNe provides us information 
about their progenitors and explosion mechanisms. \citet{2002AJ....123..745R}
made a comparative study of the distribution of peak absolute magnitudes in the
$B$-band (M$_B$) for various SNe. They found that for normal and bright SNe Ib/c, 
the mean peak M$_B$ values are $-17.61 \pm 0.74$ and $-20.26 \pm 0.33$ mag, respectively. 
The M$_B$ values were found to be $-17.56 \pm 0.38$ mag and $-19.27 \pm 0.51$ 
mag for normal and bright Type II-L SNe, while for Type II-P and IIn
SNe the M$_B$ values were found to be $-17.0 \pm 1.12$ mag and $-19.15 \pm 0.92$ mag, respectively.

In a recent study by \citet{2011MNRAS.412.1441L}, absolute magnitudes of
SNe Ibc (Type Ib, Ic and Ib/c) and II were derived using LOSS samples and the 
average absolute magnitudes (close to $R$-band as claimed by authors, 
see discussions of \citet[]{2011MNRAS.412.1441L}) were found to be
$-16.09 \pm 0.23$ mag and $-16.05 \pm 0.15$ mag for SNe Ibc and II respectively.  
In a similar study, \citet{2011ApJ...741...97D} also reported that $R$-band absolute
magnitudes of SNe Ib and Ic peak arround $-17.9 \pm 0.9$ mag and $-18.3 \pm 0.6$ mag
respectively.    

\begin{figure}
\centering
\includegraphics[scale = 0.4]{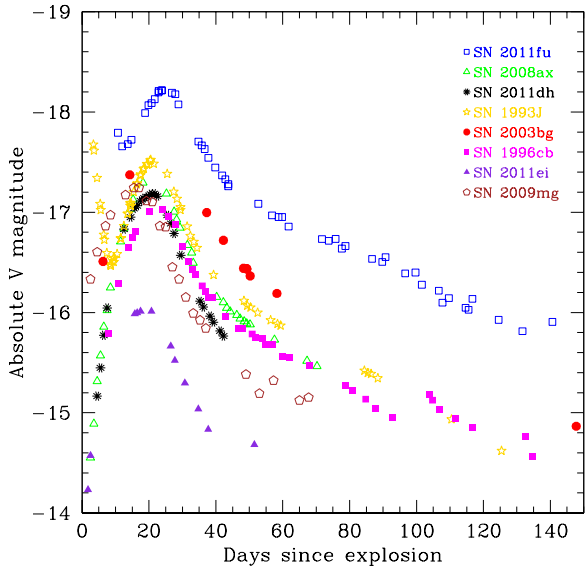}
\caption{The $M_V$ light curve of SN 2011fu is compared with those of other similar IIb 
events: SN~2011ei, 2011dh, 2009mg, 2008ax, 2003bg, 1996cb and 1993J.}
\label{fig:mabs}
\end{figure}

Fig.~\ref{fig:mabs}, shows the comparison of $V$-band absolute LC of SN~2011fu
along with seven other well-observed Type IIb SNe i.e. 1993J \citep{1994MNRAS.266L..27L},
1996cb \citep{1999AJ....117..736Q}, 2003bg \citep{2009ApJ...703.1612H}, 
2008ax \citep{2008MNRAS.389..955P}, 2009mg \citep{2012MNRAS.424.1297O}, 
2011dh \citep{2012A&A...540A..93V} and 2011ei \citep{2012arXiv1207.2152M}. 
For SN~2011fu, the distance $D$ = 77.9 $\pm$ 5.5 Mpc
has been taken from the NED\footnote{http://ned.ipac.caltech.edu/} along with total
$E(B-V)$ = 0.22 $\pm 0.11 $ mag as discussed in previous subsection.
However, all other LCs presented in the figure have been corrected for 
interstellar extinctions and distance values collected from literature. 
Fig.~\ref{fig:mabs}, illustrates that the peak M$_V$ for various Type IIb SNe has 
a range between $\sim$ $-16$ mag and $\sim$ $-18.5$ mag. 
In this distribution, SN~2011fu is the brightest from early to late epochs with a
peak absolute magnitude of M$_V$ $\sim$ -18.5 $\pm 0.24 $ mag.

\section{Bolometric light curve}

\subsection{Construction of the bolometric light curve}

The quasi-bolometric lightcurve (\emph{UBVRI}) was computed by integrating the
extinction-corrected flux\footnote{Fluxes were corrected for interstellar reddening using the 
{\sc idl} program $ccm$\_$unred.pro$ available at ASTROLIB (http://idlastro.gsfc.nasa.gov/ftp/)
by adopting $E(B-V)$ = 0.22 mag for the total (Milky Way plus in-host) reddening
and by assuming reddening law for the diffused interstellar medium ($R_v$ = 3.1).} 
in all 5-bands. The data were interpolated wherever it was necessary 
and total \emph{UBVRI} flux was integrated using a simple trapezoidal rule.

\begin{figure}
\centering
\includegraphics[scale = 0.4]{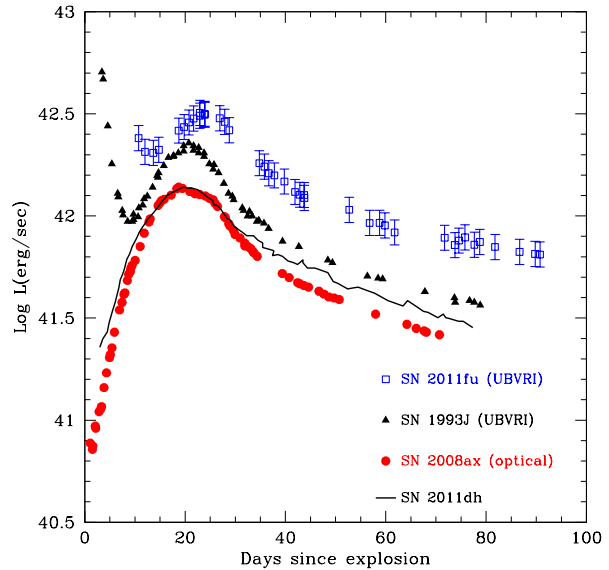}
\caption{The bolometric light curve of SN 2011fu compared with similar Type IIb
events SN 1993J \citep{1994MNRAS.266L..27L}, SN 2008ax \citep{2008MNRAS.389..955P}
and SN 2011dh (Ergon et al. 2012)}
\label{fig:mbol}
\end{figure}

In Fig.~\ref{fig:mbol}, we compare the quasi-bolometric LC of SN~2011fu  
along with other three Type IIb events i.e. 
SNe 1993J \citep{1994MNRAS.266L..27L}, 2008ax \citep{2008MNRAS.389..955P} 
and SN 2011dh \citep{2012In...preparation1}. It is obvious that the shape 
of the quasi-bolometric LC of SN~2011fu is similar to that of SN~1993J. 
However, SN~2011fu is more luminous in comparison with the other SNe 
during the observed phases. 

The un-observed part of the bolometric LC in the Infra-red was approximated
by assuming blackbody flux distributions fitted to the observed $R$- and 
$I$-band fluxes for each epoch. At first, we used the Rayleigh-Jeans 
approximation for the fluxes redward of the $I$-band, and integrated the 
flux distribution between the $I$-band central wavelength and infinity. 
This resulted in an analytic estimate for the IR (infra-red) contribution
as $L_{IR} \approx \lambda_I \cdot F_I / 3$, where $\lambda_I$ and $F_I$ are the 
$I$-band central wavelength and monochromatic flux, respectively. Second, we 
fitted a blackbody to the $R$- and $I$-band fluxes at each epoch, and 
numerically integrated the fitted blackbody flux distributions from
the $I$-band to radio wavelengths ($\sim 1$ mm). These two estimates gave
consistent results within a few percent, which convinced us that they
are more-or-less realistic estimates of the IR-contribution. Because the 
$R$-band fluxes may also be affected by the presence of $H\alpha$,
we adopted the result of the first, analytic estimate as the final result.
The comparison of the integrated \emph{UBVRI}- and IR-fluxes showed that the 
IR-contribution was $\sim 20$ percent at the earliest observed phases, 
but it increased up to $\sim 50$ percent by day $+40$ and stayed 
roughly constant after that. 

\subsection{Bolometric light curve modelling}

The bolometric light curve (see subsection 4.1) 
was fitted by the semi-analytic light curve model of 
\citet{1989ApJ...340..396A} \citep[see also][]{2009ApJ...704.1251C}.
This model assumes homologously expanding spherical ejecta having constant
opacity, and solves the photon diffusion equation taking into account the
laws of thermodynamics. This approach was first introduced by 
\citet{1980ApJ...237..541A} and \citet{1982ApJ...253..785A}, and further
extended by \citet{1989ApJ...340..396A} by taking into account the rapid 
change of the opacity due to recombination. The extended diffusion-recombination 
model was succesfully applied to describe the observed light curve of SN~1987A 
assuming realistic physical parameters \citep{1989ApJ...340..396A}.

The bolometric LC of SN~2011fu is qualitatively similar to that of
SN 1987A, because of the presence of the rapid initial decline
and the secondary bump, after which the LC settles down onto the
radioactive tail due to the $^{56}$Co-decay. This early LC decline in not unusual in 
Type IIb SNe (however, see Fig.~\ref{fig:snid} at early 
epoch where we compare the LCs of SN 2011fu with SN 1993J), and it is usually modelled by a 
two-component ejecta configuration: a dense compact core
and a more extended, lower density envelope on top of the core
\citep{2012ApJ...757...31B}. The fast, initial decline is thought to be 
due to the radiation of the cooling outer envelope (which was
initially heated by the shock wave passing through it after the explosion), 
while the secondary bump is caused by the photons diffusing slowly out 
from the inner, denser ejecta which is mainly heated from inside by the 
radioactive decay of $^{56}$Ni $\rightarrow$ $^{56}$Co $\rightarrow$ $^{56}$Fe. 
After the secondary maximum, the decline of the LC is faster than the rate of 
the radioactive decay, which may be due to a recombination front moving inward 
into the ejecta, similar to the condition at the end of the plateau phase in 
Type II-P SNe.

\begin{table*}
\caption{Log of parameters derived from bolometric light curve modelling, discussed in section 4.2.}
\label{tab:lcmodel}
\begin{tabular}{lccl}
\hline \hline
Parameters & He-core & H-envelope & remarks\\
\hline
$R_{prog}$ (cm) & $2 \times 10^{11}$ & $1 \times 10^{13}$ & progenitor radius\\
$M_{ej}$ ($M_\odot$) & $ 1.1$ & $0.1$ & ejecta mass \\
$\kappa_T$ (cm$^2$g$^{-1}$) & $0.24$ & $0.4$ & Thompson scattering opacity \\
$M_{Ni}$ ($M_\odot$) & $0.21$ & $-$  & initial nickel mass \\
$E_{kin}$ (10$^{51}$ erg) & $2.4$ & $0.25$ & ejecta kinetic energy \\
$E_{th}(0)$ (10$^{51}$ erg) & $1.0$  & $0.3$ & ejecta initial thermal energy \\
\hline
\end{tabular}
\end{table*}

In order to simulate this kind of LC behavior, we slightly modified the 
original diffusion-recombination model of \citet{1989ApJ...340..396A}.
Instead of having a H-rich, one-component ejecta, we added an extended, 
low-density, pure H envelope on top of a denser, He-rich core. Following 
\citet{1989ApJ...340..396A}, we also assumed that the opacity is due 
to only Thompson-scattering, and it is constant in both the envelope 
and the core. Because the envelope was thought to contain only 
H, $\kappa = 0.4$ cm$^2$ g$^{-1}$ was selected as the Thompson-scattering 
opacity for this layer, while $\kappa = 0.24$ cm$^2$ g$^{-1}$ was applied 
for the inner region to reflect its higher He/H ratio. 

The system of differential equations given by \citet{1989ApJ...340..396A} 
were then solved by simple numerical integration (assuming a short, 
$\Delta t = 1$ s timestep which was found small enough to get a
reasonable and stable solution). Because the photon diffusion 
timescale is much lower in the envelope than in the core, the 
contribution of the two regions to the overall LC is well separated: 
during the first few days the radiation from the outer, adiabatically 
cooling envelope dominates the LC, while after that only the photons 
diffusing out from the centrally heated inner core contribute. Thus, 
the sum of these two processes determines the final shape of the LC.

\begin{figure}
\centering
\includegraphics[scale =0.45]{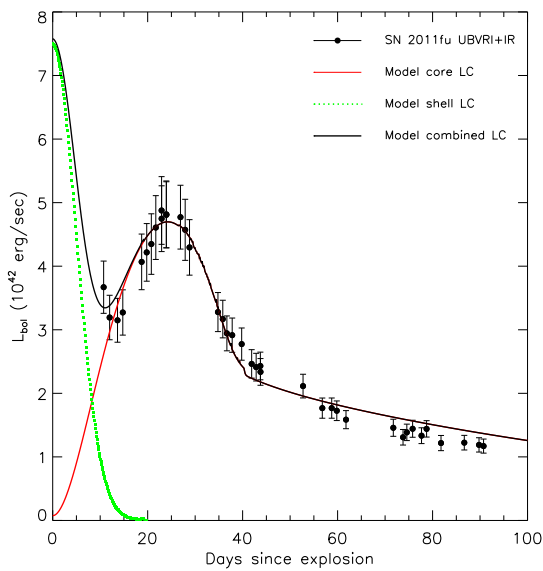}
\caption{Comparison of the observed bolometric LC (dots) with the best-fit
two-component diffusion-recombination model. The red and green curves 
show the contribution from the He-rich core and the low-mass H-envelope,
respectively, while the black line gives the combined LC.}
\label{fig:lcmodel}
\end{figure}

Because of the relatively large number of free parameters, we
have not attempted a formal $\chi^2$ minimization while fitting
the model to the observations. Instead, we searched for a 
qualitative agreement between the computed and observed bolometric LCs.
The parameters of our final, best-fit-by-eye model are collected in
Table~\ref{tab:lcmodel}, while the LCs are plotted in Fig.~\ref{fig:lcmodel}.

It is seen that the best-fit model consists of a dense, $1$ $M_\odot$ 
He-rich core and a more extended, low-mass ($0.1$ $M_\odot$) H-envelope.
This is very similar to the progenitor configuration found by
\citet{2012ApJ...757...31B} when modelling the LC of another Type IIb event, 
SN~2011dh, although they assumed a more massive ($\sim 3$ $M_\odot$) He-core. 
Nevertheless, it was concluded by \citet{2012ApJ...757...31B} and confirmed 
by the present study that the secondary bump is entirely due to radiation 
coming from the dense inner core of the ejecta, and the outer extended 
envelope is only responsible for the initial fast decline of the LC. 
The estimated ejecta mass for SN~2011fu, $\sim 1.1 $ $M_\odot$ is consistent 
with the observed rise time ($\sim 24$ days) to the secondary maximum of 
the LC \citep[see Eq.10 of][]{2012ApJ...746..121C}. The parameters in
Table~\ref{tab:lcmodel} are also qualitatively similar to the ones derived by
\citet{1995ApJ...449L..51Y} for modelling the LC of SN~1993J.

There are a number of caveats in the simple diffusion-recombination model used
above, which naturally limit the accuracy of the derived physical parameters.
The most obvious limitation is the assumption of constant opacity in the ejecta.
The pre-selected density and temperature profiles in the ejecta (assumed as 
exponential functions) are also strong simplifications, but they enable the 
approximate, semi-analytic treatment of the complex problem of radiative diffusion, 
as shown by \citet{1989ApJ...340..396A}. Thus, the parameters in 
Table~\ref{tab:lcmodel} can be considered only as order-of-magnitude estimates, 
which could be significantly improved by more sophisticated modelling 
codes \citep[e.g.][]{2012ApJ...757...31B}. 

\begin{figure*}
\centering
\includegraphics[scale = 0.7]{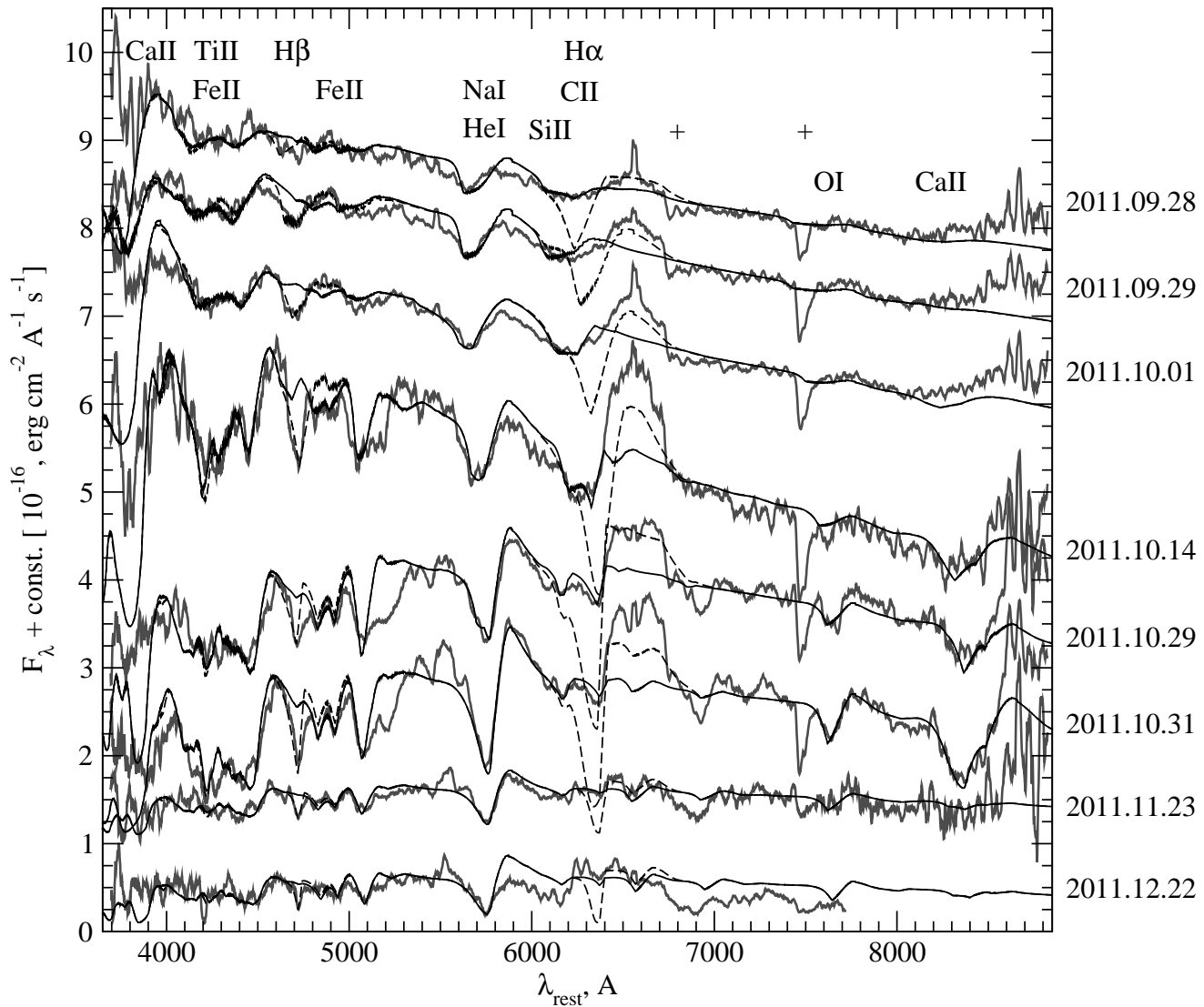}
\caption{The evolution of the SN 2011fu spectra (grey thick curves, smoothed by a
20\AA{}-wide window function) overplotted with \texttt{SYNOW} models.
The main models are shown by the solid black line. The models
with $H\beta$ fitting are shown as dashed black line. The most conspicuous
ions are marked. Atmospheric lines are marked with ``+''.}
\label{Spectra}
\end{figure*}

\section{Spectral analysis}

Properties of the SN~2011fu ejecta were investigated with the multi-parametric
resonance scattering code \texttt{SYNOW} \citep{1997ApJ...481L..89F}
\citep[see also][]{2002ApJ...566.1005B, 2005ASPC..342..351B, 2006A&A...450..305E}.
The evolution of temperature and velocities of layers were traced through
several months of spectral observations.
The \texttt{SYNOW} code is based on several assumptions: spherical symmetry;
homologous expansion of layers ($v \sim r$); sharp photosphere producing a
blackbody spectrum and associated with a shock wave at early stages.

\subsection{Comparision between observed and synthetic spectra}
In the photospheric phase the spectral lines with P Cygni profiles are
formed by resonance scattering in a shell above the optically thick 
photosphere which produces the continuum \citep[see][]{2001astro.ph.11573B}. 
On the other hand, during the nebular phase the ejecta is transparent (optically
thin) in the optical wavelength range. In this case the spectrum is dominated by strong
emission features including forbidden lines. Each of these two phases of SN evolution 
can be explained with individual approximations and the modelling of the observed 
spectra should be made with different synthetic codes. There is no sharp boundary
between these two phases. No strong transition to the nebular phase with conspicuous 
emission features can be seen in the observed spectra of SN~2011fu (Fig.~\ref{Spectra}). 
The shape of lines remains the P Cyg profile, which suggests that they are formed
by resonance scattering, as assumed in \texttt{SYNOW}. Thus, we modelled all spectra
of SN~2011fu with this code.   
Before modelling, all spectra have been corrected for redshift (see Sect.~\ref{sec:phot}).

The strong emission component of the $H\alpha$ line (probably with the C\,{\sc ii} and
Si\,{\sc ii} contamination) can not be fully fitted in the terms of the \texttt{SYNOW} 
code. We focused primarily on the absorption parts of P Cyg line profiles which provide 
information about the expansion velocities of different line-forming layers. 
The \texttt{SYNOW} code allows the usage of different optical depth (i.e. density)
profiles. Two of them are the exponential profile with the parameter of e-folding velocity
``$v_e$''  ($\tau \propto exp (-v/v_e)$) which can be adjusted for each ion, and the 
power-law profile with the index ``$n$'' ($\tau \propto v^{-n}$ ) which is applied
to all ions in the model. We checked both cases and found that the exponential law is 
more suitable for our spectra. The original paper of the \texttt{SYNOW} developers 
and further studies showed a possibility of spectral features which can be detached or 
undetached from the photosphere. These two configurations produce different shapes for 
the line profiles, which was described in the paper by \citet{2008AstBu..63..228S}.

The first three observed spectra are separated by only one and two days. That is why 
they can be modelled by similar sets of parameters (see Table~\ref{Table}). Even the 
spectrum obtained on Oct 14 has a similar continuum slope ($T_{bb}\approx6500-6700K$).
To verify the pseudo-photospheric temperature derived by \texttt{SYNOW} modelling, 
we also evaluated the colour temperature ($T_{col}$) of the SN using the models 
of \citet{2005A&A...437..667D} and \citet{2009ApJ...701..200B}. We used the $B-V$ 
colours for those epochs where spectra were available, and then estimated the 
temperature from the corresponding $B-V$ colour. Both of these temperature estimates 
seem to be consistent except for the spectra taken on Oct 14 and Dec 22, 2011.

\subsection{Velocity of the pseudo-photosphere}
\label{sec:Vph}

The velocity of the pseudo-photosphere (an optically thick layer, the surface of last
scattering for continuum photons) can be located by velocities of heavy elements such 
as Fe\,{\sc ii} and Ti\,{\sc ii}, which may produce optically thin spectral features. 
However, during the very early phases these features are very weak and blended. 
Therefore, fitting the first three spectra by these ions gives a wide range of 
possible photospheric velocities, extending from 13~000 to 19~000 km~s$^{-1}$. 
The most prominent, narrow absorption feature in these spectra is the feature near 
5650\AA{} produced by He\,{\sc i} (which may be blended with Na\,{\sc i} D). This 
feature is useful to better constrain the velocity at the pseudo-photosphere, and 
decrease the uncertainty of this parameter at the earliest phases. All velocities 
derived this way are shown in the $V_{phot}$ column of Table~\ref{Table}.

\subsection{Hydrogen and the 6200\AA{} absorption feature} \label{sec:Hydrogen}

The wide absorption feature near 6200\AA{} can be fitted with the help of
a high-velocity H-layer (up to $V\sim$20~000~km~s$^{-1}$) which may be detached from
the pseudo-photosphere. On the other hand, fitting the emission peak of H$\alpha$ with 
\texttt{SYNOW} needs lower velocities, but those models cannot reproduce the 
absorption profile (see Fig.~\ref{Spectra}). In the $V(H\alpha)$ column of 
Table~\ref{Table} we list the results from the latter, more conservative solution.

The broad absorption at 6200\AA{} can be also explained with the presence of the
C\,{\sc ii} ion having a high-velocity almost identical to that of H$\alpha$. 
Moreover, C\,{\sc ii} also produces a small feature near 4400\AA{}. 
This feature can constrain the reference optical depth ($\tau$) for ionized carbon. 
But the contamination from heavy elements in the blue region makes the
fitting of the C\,{\sc ii} 4400\AA{} feature uncertain. Thus, the presence of
carbon cannot be confirmed from these spectra.

It is also possible to explain the 6200\AA{} feature by singly ionized silicon. 
In this case the velocity of Si\,{\sc ii} must be very low. On the other hand,
it is expected that the velocity of Si\,{\sc ii} should be equal or only slightly 
higher than the photospheric velocity. It turned out that 
only the blue wing of this wide feature can be fitted by Si\,{\sc ii}. Although
the small absorption near 5880 \AA{} might be explained by the presence of
Si\,{\sc ii}, the observed shape of the 6200\AA{} feature does not confirm this
hypothesis. 

\begin{figure*}
\begin{centering}
\includegraphics[width = 12cm, height = 10cm]{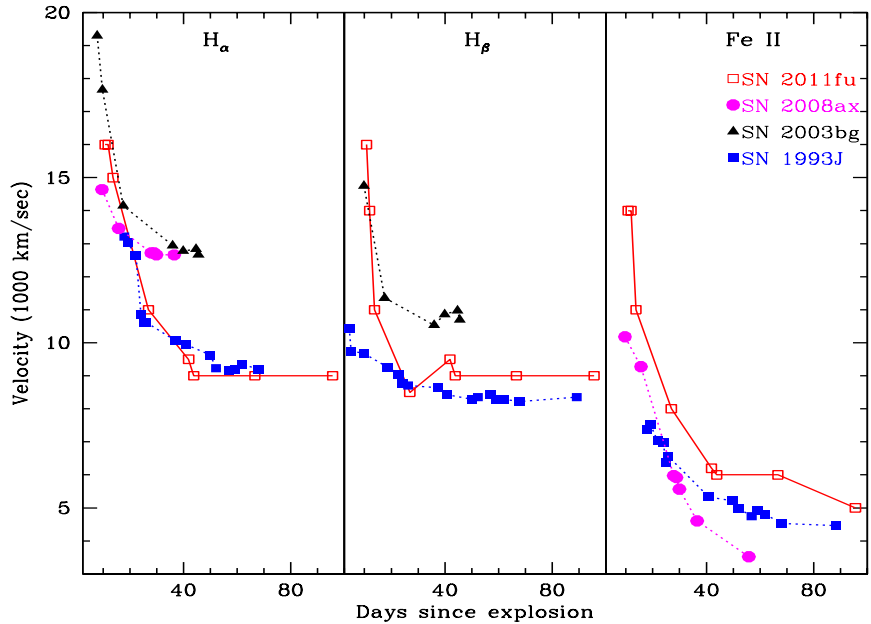}
\caption{The evolution $H\alpha$, $H\beta$ and Fe\,{\sc ii} line velocities by 
fitting \texttt{SYNOW} model (see Table \ref{Table}). Photospheric velocities for SN~2011fu, 
2003bg \citep{2009ApJ...703.1612H}, 1993J \citep{1994MNRAS.266L..27L, 1995A&AS..110..513B} 
and 2008ax \citep{2008MNRAS.389..955P} are shown. The symbols of SN~2011fu are connected 
with lines, those of other SNe with dotted lines.}
\label{Vel}
\end{centering}
\end{figure*}

\begin{table}
\caption{Velocities of the pseudo-photosphere, $H\alpha$ and $H\beta$
for different epochs of SN~2011fu evolution, derived with \texttt{SYNOW}.
We assumed that the photospheric velocity ($V_{phot}$) is equal to the velocity of
Fe\,{\sc ii}. All velocities are given in km~s$^{-1}$.
$T_{bb}$ is the blackbody temperature of the pseudo-photosphere in Kelvin degrees.
The colour temperature ($T_{col}$) derived from effective temperature $-$ colour
relations \citep[see][]{2005A&A...437..667D, 2009ApJ...701..200B} is given in last column.}
  \label{Table}
  \begin{tabular}{cccccc}
  \hline \hline
UT Date       & $V_{phot}$     & $V(H\alpha)$ &  $V(H\beta)$ &  $T_{bb}$ & $T_{col}$ \\
(yyyy/mm/dd)  & ($Fe\,II$)     &              &              &           &           \\   \hline
2011/09/28 &14000   &16000  &16000    & 6700 & 6952           \\
2011/09/29 &14000   &16000  &14000    & 6500 & 6476           \\
2011/10/01 &11000   &15000  &11000    & 6500 & 6052           \\
2011/10/14 &8000    &11000  & 8500    & 6700 & 5791           \\
2011/10/29 &6200    & 9500  & 9500    & 5000 & 4698           \\
2011/10/31 &6000    & 9000  & 9000    & 5000 & 4795           \\
2011/11/23 &6000    & 9000  & 9000    & 5000 & 5094           \\
2011/12/22 &5000    & 9000  & 9000    & 5000 & 5718           \\
  \hline
  \end{tabular}
  \end{table}

In order to look for other possibilities, we also checked different blends of 
H, Si\,{\sc ii}, C\,{\sc ii} and some other ions with different 
velocities (assuming undetached as well as detached line formation) in our models. 
At the early phases the range of derived velocities turned out to be wide due to 
the lack of observable spectral features formed close to the photosphere, as 
discussed above. At the late phases, the wide absorption near 6200\AA{} splitted 
into at least three separate features (6100\AA{}, 6200\AA{}, 6350\AA{}). 
These features might be explained as a line formation effect for H$\alpha$ in layers 
with different velocities or the appearance of blending due to ions mentioned above. 
Unfortunately, no firm conclusion can be drawn based on the simple parametric models 
that \texttt{SYNOW} can produce. 

The deep absorption near 4700\AA{} can be naturally explained identifying it as
the $H\beta$ line. We fitted this line and the $H\alpha$ line independently 
because they cannot be modeled by the same set of parameters: $\tau$, 
$v_{phot}$ and $v_e$ \citep[see also][]{2007ApJ...666.1093Q}.
From Table~\ref{Table} it is visible that for the spectra obtained before Oct 29 
the fitting of $H\alpha$ needs higher velocities than the fitting of $H\beta$.
Although both the $H\alpha$ and $H\beta$ velocities declined in time, the
formation of the absorption component of $H\alpha$ stayed at higher
velocities than for $H\beta$. This may suggest that $H\alpha$ remained optically
thick for a longer time than $H\beta$ in the expanding, diluting H-rich
envelope.

\subsection{Other ions}
\label{sec:oth_ions}
To fit the main features in the spectra, we included the following elements 
and ions in \texttt{SYNOW}: 
H, Fe\,{\sc ii}, Ti\,{\sc ii}, Na\,{\sc i}, Si\,{\sc ii}, O\,{\sc i} and Ca\,{\sc ii}. 
Some models were also computed containing the following elements 
and ions as alternatives: C\,{\sc ii}, He\,{\sc i}, Fe\,{\sc i},  
Ti\,{\sc i}, Sc\,{\sc ii}, Mg\,{\sc i} and the consistency of these ions were  
cross-examined with \citet{1999ApJS..121..233H}.
The heavier atoms/ions should have velocities close to $v_{phot}$ but the
lighter ones may be detached due to e.g. stratification of elements in the ejecta.

In the followings, we show some possibilities to explain these features 
in the observed spectra. Weak features near 4810\AA{} and 6370\AA{} can be 
explained by the presence of low-velocity He\,{\sc i}. He can also be found 
as a blend with Na\,{\sc i} in the deep absorption near 5650\AA{} mentioned above,
and as a blend with Ti\,{\sc ii} and Fe\,{\sc ii} near 4300\AA{}. However, the velocity 
of He\,{\sc i} may be higher than the photospheric velocity. Even in this case, 
the presence of He can explain all these features. 

A small absorption in the blue wing of $H\alpha$ at 6630\AA{} could be modeled 
by He\,{\sc i} or low-velocity C\,{\sc ii}.  
The feature near 5050\AA{} could be fitted with Mg\,{\sc i} as well. The Ca\,{\sc ii} H+K 
feature cannot be fitted well around 3730\AA{}, because this regime is at the
blue end of our observed spectra and all of them are very noisy at these wavelengths. 
But the absorption feature near 8400\AA{} is compatible with the Ca\,{\sc ii} IR-triplet.

\subsection{Results of spectral modelling}
\label{sec:SYNOW_results}

Almost all spectral features are described well with elements and ions
which are usually applied in the case of Type IIb SNe. However, the strong
emission of $H\alpha$ dominating during intermediate epochs can not be fully
fitted with the models applied.

The fitting of redder and especially bluer parts has some uncertainties
due to strong blending of metal lines such as Ca\,{\sc ii}, Ti\,{\sc ii}, Fe\,{\sc ii} 
and others. Even without precise modelling, all spectral sequences can be divided
into two groups: the first four spectra (up to Oct 14) which are fitted with models with the
blackbody temperature  $T_{bb} \approx 6500 $ - $ 6700$ K and the following four spectra 
with $T_{bb} \approx 5000$ K. 

Generally, the modelling of the SN 2011fu spectra shows the decline of the
photospheric velocities up to $\sim40$ days after the explosion (see Fig.~\ref{Vel}).
Then, all velocities remain approximately at the same, stable level. This
behaviour was also described in previous works on CCSNe 
\citep{2002ApJ...566.1005B, 2007ApJ...666.1093Q, 2010AstBu..65..132M}.
In Fig.~\ref{Vel} we plot the velocities of H$\alpha$, H$\beta$ and 
Fe\,{\sc ii} for SN 2011fu, SN 2008ax, SN 2003bg and SN 1993J, illustrating this
effect.

\section{Metallicity-Brightness comparision of host galaxies}
   
In several earlier studies of CCSNe hosts, it has been already mentioned that
various SNe subtypes occur in different environments 
\citep[see][]{2008ApJ...673..999P, 2010MNRAS.407.2660A, 2010ApJ...721..777A, 
2011ApJ...731L...4M, 2012ApJ...759..107K, 2012ApJ...758..132S}. 
Metallicity is a key factor in all these studies. Recent studies by \citet{2010ApJ...721..777A} 
and \citet{2008ApJ...673..999P} found that SN Ib/c host galaxies are metal-rich as compared 
to SN II hosts. 
\citet{2011ApJ...731L...4M} found that SNe Ic are more metal-rich (up to 0.20 dex) 
than SNe Ib. In a similar study on SNe Ib/c locations \citet{2011A&A...530A..95L} found  
a smaller gap between the two metallicities 
(the environment of SNe Ic is richer by $\sim 0.08$ dex than Ib). In a recent study with a 
different approach (using local emission-line for metallicity estimates), where 74 H\,{\sc ii} regions 
in CCSNe hosts were analyzed, \citet{2010MNRAS.407.2660A} did not find difference between 
the metallicities of these two environments. 

Type IIb host galaxies have been claimed to be more metal-poor than those of SNe Ib or Ic 
\citep[see][]{2010ApJ...721..777A, 2012ApJ...759..107K}, although in another recent study 
which is based on the SN sample from untargeted searches (although with a rather small 
sample of 8 SNe IIb) \citet{2012ApJ...758..132S} found that the median metallicity of both 
SNe Ib and IIb host galaxies is very similar. 

\begin{figure}
\begin{centering}
\includegraphics[scale = 0.45]{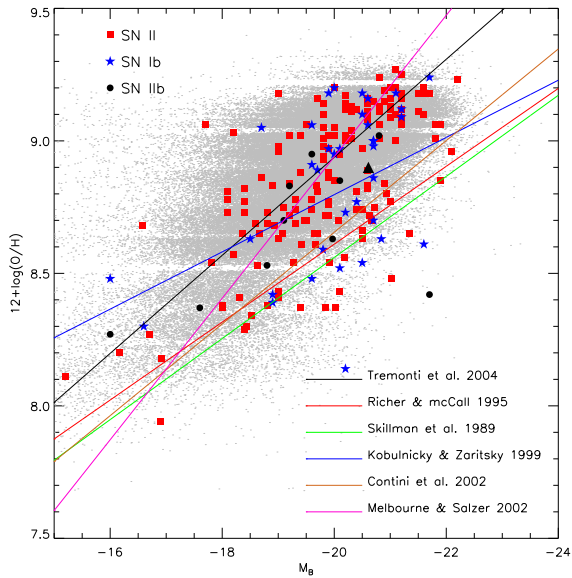}
\caption{Metallicity-luminosity relation for various types of SNe host galaxies. The
tiny dots belong to all galaxies used by \citet{2008ApJ...673..999P} 
\citep[This catalog is based on SDSS DR4][database]{2006ApJS..162...38A}.
Red squares are Type II, stars are Type Ib/c and black dots are Type
IIb SNe, respectively. The analytic relations collected from several papers 
(see text) are also over-plotted. SN 2011fu host metallicity is denoted by a black triangle.}
\label{metal}
\end{centering}
\end{figure}

In an attempt to understand the metallicity scenario for the SN 2011fu host
galaxy, we collected the latest sample of metallicity data for hosts of CCSNe, 
and their absolute magnitudes in the $B$-band from literatures 
\citep{2011A&A...530A..95L, 2011ApJ...731L...4M, 2012arXiv1205.2338S} and 
available online \footnote{www.astro.princeton.edu/$\sim$jprieto/snhosts/}. 
In Fig.~\ref{metal}, the data for these host galaxies  
(36 for Ib, 15 for IIb and 167 for remaining II SNe) were then overplotted to 
the sample containing all star forming galaxies from SDSS DR4
\citep[This sample was taken from][]{2008ApJ...673..999P}. The relations
between metallicity and $M_{\rm B}$ for galaxies from several papers are also over-plotted 
\citep{1989ApJ...347..875S, 1995ApJ...445..642R, 1999ApJ...511..118K, 
2002MNRAS.330...75C, 2002AJ....123.2302M, 2004ApJ...613..898T}.

We estimated the metallicity of the host of SN 2011fu using the relation given
by \citet[][]{2002ApJ...581.1019G} (see their equation 6). We considered 
M$_{\rm B}$ = -20.62 mag for UGC 01626 from HyperLeda. The calculated 
$\log(O/H)+12$ for UGC 01626 is 8.90$^{+ 0.10}_{- 0.06}$. 
This value is slightly higher than 
$\log(O/H)+12$ = 8.55 \citep{2010ApJ...721..777A} and $\log(O/H)+12$ = 8.44
\citep{2012ApJ...758..132S} for SN type II sample.
Our analysis using most updated sample of absolute magnitudes and metallicities of CCSNe 
host galaxies also supports the results described in \citet{2012ApJ...758..132S}.      
However, it is noticeable that methods used to determine metallicities are based on 
statistical samples, affected by incompleteness of the sample and should be used with caution.

\section{Conclusions}
We present a comprehensive \emph{UBVRI} photometeric and low-resolution spectroscopic 
monitoring of the Type IIb SN 2011fu. To date, only a handful of SNe belonging to 
this class have been observed and studied in detail. 

To the best of our knowledge, our photometric and spectroscopic observations described 
here are the earliest ones reported for this event. The early photometric observations 
strongly suggest the presence of the early-time decline of the light curve (which is 
thought to be related to the shock break-out phase) as seen in case of SN~1993J. 
The early-time LC decay rate ($\alpha$$_1$) of this SN is slower than that derived for 
SN~1993J in all the bands. The rising rates between the LC valley to peak ($\alpha$$_2$) 
observed in SN~2011fu is also somewhat slower than in SN~1993J. However, 
the post peak LC decay rate ($\alpha$$_3$) are similar in the two events. 

The colour evolutions of SN 2011fu were studied using our \emph{UBVRI} band observations. 
Our data showed that during the very early phases the $B-V$ colour was very similar to that 
in SN~1993J. A similar trend has been found in the $V-R$ and $V-I$ colours as well. 
The evolution of these three colours after +40 days were also similar to those seen in 
other CCSNe. The $V$-band absolute magnitudes of a sample of 8 Type IIb SNe were compared 
after applying proper extinction corrections and taking into account distances collected 
from the literature. In this sample, SN 2011fu seems to be the most luminous event. 
However, the peak $V$-band absolute magnitude of SN~2011fu is not an outlier when is 
compared to the peak brightness of CCSNe of other types. 

The quasi-bolometric LC of SN 2011fu was assembled using our \emph{UBVRI}
data and accounting for the IR contribution as specified in Section 4. 
The comparison of these data with other known Type IIb SNe also shows 
that SN~2011fu is the brightest Type IIb SN in the sample.
The bolometric LC was modeled by applying a semi-analytical model of 
\citet{1989ApJ...340..396A}. This model suggests a 1.1 $M_\odot$ He-rich core 
and an extended, low-mass ($\sim 0.1$ $M_\odot$) H-envelope as the progenitor 
of SN~2011fu, similar to that of SN~2011dh. However, the progenitor radius of 
SN~2011fu ($\sim$ $ 1 \times 10^{13}$ cm) turned out to be smaller than that of 
SN~1993J ($\sim$ $4 \times 10^{13}$ cm) \citep{1994ApJ...429..300W}. 
The ejected nickel mass for SN 2011fu was $\sim 0.21$ $M_\odot$, higher than 
that of SN~1993J (0.07 $-$ 0.11 $M_\odot$). 

The spectra of SN~2011fu taken at eight epochs were analyzed using the multi-parameter 
resonance scattering code \texttt{SYNOW}. The derived parameters describe the evolution 
of the velocities related to various atoms/ions and the variation of the blackbody 
temperature of the pseudo-photosphere. The photospheric velocities at the early epochs 
were higher than that of other Type IIb SNe. The pseudo photospheric temperatures 
were found to be between 6700 K and  5000 K, decreasing from initial to later phases. 
The temperatures from \texttt{SYNOW} were also checked by comparing them with colour 
temperatures calculated from $B-V$ vs $T_{eff}$ relations 
\citep{2009ApJ...701..200B, 2005A&A...437..667D}. These different temperature 
estimates were found being consistent. The appearance of the main observed spectral 
features were also successfully modeled with \texttt{SYNOW} by assuming H, He\,{\sc i} and
various metals (mostly Fe\,{\sc ii}, Ti\,{\sc ii} and Ca\,{\sc ii}), which are
typical of CCSNe spectra. 
The estimated value of the metallicity of the host galaxy of SN~2011fu is 8.90$^{+ 0.10}_{- 0.06}$
similar to those for other Type IIb SNe.

\section*{Acknowledgements}
We thank the referee, Andrea Pastorello for a critical reading of
the paper and several useful comments and suggestions, which
greatly improved the scientific content of the manuscript. 
We are also thankful to the observers at the Aryabhatta Research Institute of 
observational sciencES (ARIES) who provided their valuable time and support for 
the observations of this event. We are thankful to the staffs of 1.3m DFOT and HCT  
for their kind cooperation. SBP acknowledge the support of Indo-Russian
(DST-RFBR) project No. INT/RFBR/P-100 for this work. This work was also supported by the 
following grants: The RFBR grant 11-02-12696-IND-a; the program no. 
17 ``Active processes in galactic and extragalactic objects'' of the Department 
of Physical Sciences of the Russian Academy of Sciences and grants 
No. 14.B37.21.0251 and No. 14.A18.21.1179 from FTP of the RF Ministry
of education and science. The work of JV and AO has been supported by Hungarian 
OTKA Grant K76816. BK is thankful to Eswaraiah, C. for his help during writing 
of this article. BK also acknowledge the support of Actions de Recherche Concert\'ees (ARC).
This research has made use of the NASA/IPAC Extragalactic Database (NED) which
is operated by the Jet Propulsion Laboratory, California Institute of Technology,
under contract with the National Aeronautics and Space Administration. 
We acknowledge the usage of the HyperLeda database (http://leda.univ-lyon1.fr).
\label{lastpage}
\bibliography{./sn2011fu_ref}

\end{document}